\documentclass{elsart}

 \usepackage{graphicx}

\usepackage{amssymb}

\begin{document}

\begin{frontmatter}



\title{Recent results in the BFKL theory\thanksref{FFK}}
\thanks[FFK]{This talk is based on results obtained in collaboration
with V.S.~Fadin, R.~Fiore and M.I.~Kotsky.}

\author{Alessandro Papa}

\address{Dipartimento di Fisica, Universit\`a della Calabria \\
\& Istituto Nazionale di Fisica Nucleare, Gruppo Collegato di Cosenza \\
I-87036 Arcavacata di Rende, Cosenza, Italy}

\begin{abstract}
The Balitsky-Fadin-Kuraev-Lipatov (BFKL) approach for the cross sections at 
high energy $\sqrt s$ in perturbative QCD is briefly reviewed. The role of 
gluon Reggeization in the derivation of the BFKL equation and its compatibility
with $s$-channel unitarity (``bootstrap'') are discussed. 
\end{abstract}

\end{frontmatter}

\section{Gluon Reggeization in perturbative QCD}
\label{Reggeization}

The key role in the derivation of the BFKL equation~\cite{BFKL} for the
processes at high energy $\sqrt s$ in perturbative QCD is played by the gluon 
Reggeization. ``Reggeization'' of a given elementary particle usually means that
the amplitude of a scattering process with exchange of the quantum numbers 
of that particle in the $t$-channel goes like $s^{j(t)}$ in the 
Regge limit $s\gg |t|$. The function $j(t)$ is called ``Regge trajectory''
of the given particle and takes the value of the spin of that particle when 
$t$ is equal to its squared mass. In perturbative QCD, the notion of gluon 
Reggeization is used in a stronger sense. It means not only that a Reggeon 
exists with the quantum numbers of the gluon, negative signature and with a 
trajectory $j(t) = 1 + \omega(t)$ passing through 1 at $t=0$, 
but also that this Reggeon gives the leading contribution in each order of 
perturbation theory to the amplitude of processes with large $s$ 
and fixed (i.e. not growing with $s$) squared momentum transfer $t$.

\begin{figure}[tb]
\begin{minipage}{60mm}
\begin{center}
\includegraphics*[width=4cm]{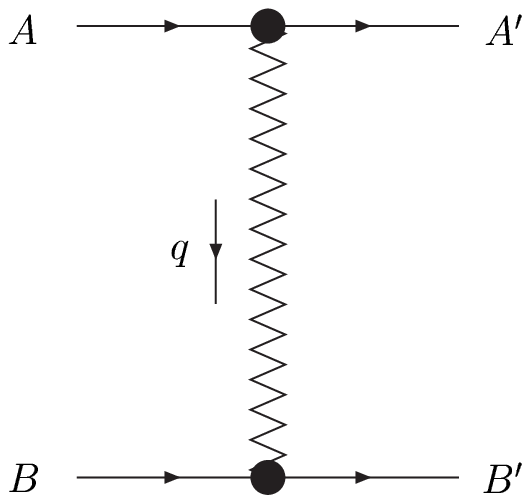}
\end{center}
\end{minipage}
\begin{minipage}{60mm}
\begin{center}
\includegraphics*[width=5.1cm]{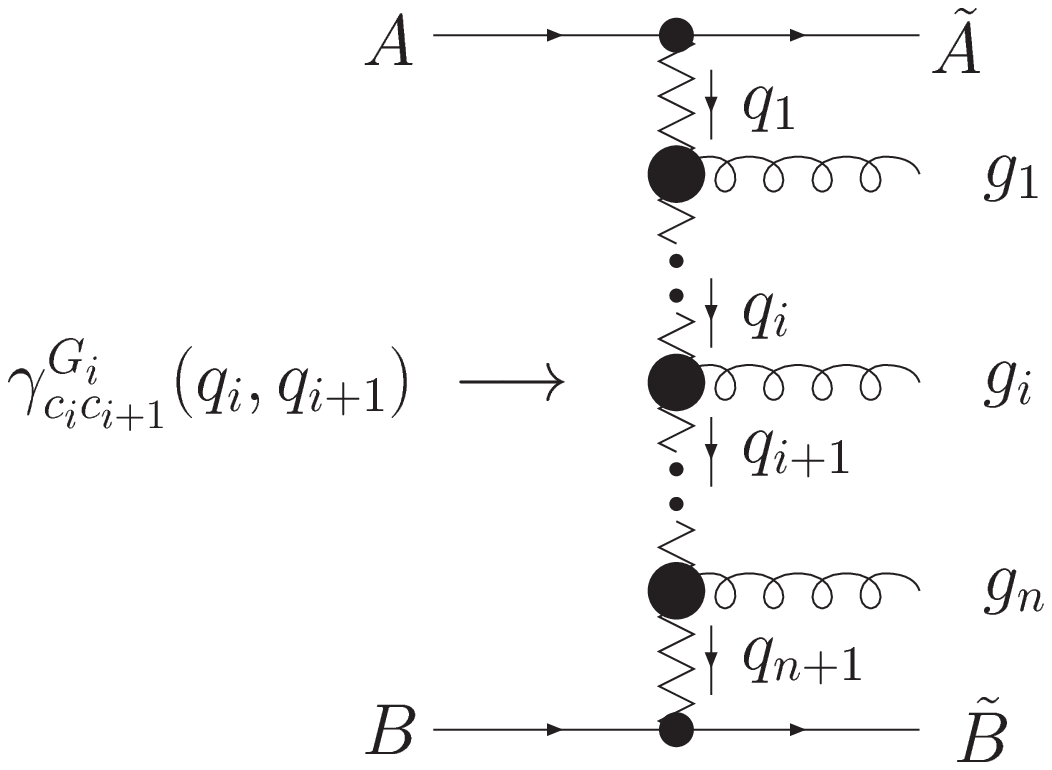}
\end{center}
\end{minipage}

\vspace{0.5cm}
\noindent {\scriptsize
Fig.~1. (Left) Diagrammatical representation of $(A_8^-)^{A^\prime B^\prime}_{AB}$.
The zig-zag line is the Reggeized gluon, the black blobs the PPR effective 
vertices.} 

\vspace{0.3cm}
\noindent {\scriptsize
Fig.~2. (Right) Diagrammatical representation of the production amplitude
$A^{\tilde A  \tilde B +n}_{AB}$ in the LLA.}
\vspace{0.5cm}

\end{figure}

To be definite, let us consider the elastic process $A + B \longrightarrow A^\prime 
+ B^\prime $ with exchange of gluon quantum numbers in the $t$-channel, i.e.
for octet color representation in the $t$-channel and negative signature
(see Fig.~1). Gluon Reggeization means that, in the Regge kinematical region 
$s \simeq - u \rightarrow \infty$, $t$ fixed (i.e. not growing with $s$), 
the amplitude of this process takes the form
\begin{equation}
\left({A_8^-}\right)^{A^\prime B^\prime}_{AB} = 
\Gamma^c_{A^\prime A}\:\left[\left({-s\over -t}\right)^{j(t)}-\left({s\over 
-t}\right)^{j(t)}\right]\:\Gamma^c_{B^\prime B}\;.
\label{elast_ampl_octet}
\end{equation}
Here $c$ is a color index and $\Gamma^c_{P^\prime P}$ are the 
particle-particle-Reggeon (PPR) vertices, not depending on $s$. They can be
written as $\Gamma^{c}_{P^\prime P} = g\langle P^\prime| T^c| P \rangle 
\Gamma_{P^\prime P}$, where $g$ is the QCD coupling constant and $T^c$ are the 
color group generators in the fundamental (adjoint) representation for quarks 
(gluons). This form of the amplitude has been proved rigorously~\cite{BFL79}
to all orders of perturbation theory in the leading-logarithmic approximation (LLA),
which means resummation of the terms $\alpha_s^n \ln^n s$. In this approximation
$\Gamma_{P^\prime P}$ is given simply by 
$\delta_{{\lambda_{P^\prime}}{\lambda_P}}$, where $\lambda_P$ is the helicity 
of the particle $P$, and the (deviation from 1 of the) Reggeized gluon
trajectory enters with 1-loop accuracy~\cite{Lip76},
$$
\omega^{(1)}(t) = {g^2 t\over {(2{\pi})}^{D-1}}\frac{N}{2}
\int{d^{D-2}k_\perp\over k_\perp^2{(q-k)}_{\perp}^2}
=-\frac{g^2 N \Gamma(1-\epsilon)}{(4\pi)^{D/2}} 
\frac{\Gamma^2(\epsilon)}{\Gamma(2\epsilon)}(-q_\perp^2)^\epsilon\;.
$$
Here $D=4+2 \epsilon$ has been introduced in order to regularize the infrared 
divergences and the integration is performed in the space transverse to the 
momenta of the initial colliding particles\footnote{In the following, since the 
transverse component of any momentum is obviously space-like, the notation 
$p_\perp^2 = - \vec {p}^{\:2}$ will be also used.}. 
In the NLA, which means resummation of the terms $\alpha_s^{n+1} \ln^n s$, 
the form~(\ref{elast_ampl_octet}) has been checked in the first three orders of 
perturbation theory~\cite{FFKQ95-96} and is only assumed to be valid to all orders. 
In this approximation, the NLA contribution to the PPR vertices takes the form
$\Gamma_{P^\prime P} = 
\delta_{\lambda_{P^\prime}\lambda_P}\Gamma^{(+)}_{PP}+
\delta_{\lambda_{P^\prime},-\lambda_P}\Gamma^{(-)}_{PP}$, where a 
helicity non-conserving term appeared, and the Reggeized gluon trajectory enters
with 2-loop accuracy.

\begin{figure}[tb]
\begin{minipage}{40mm}
\includegraphics*[width=4cm]{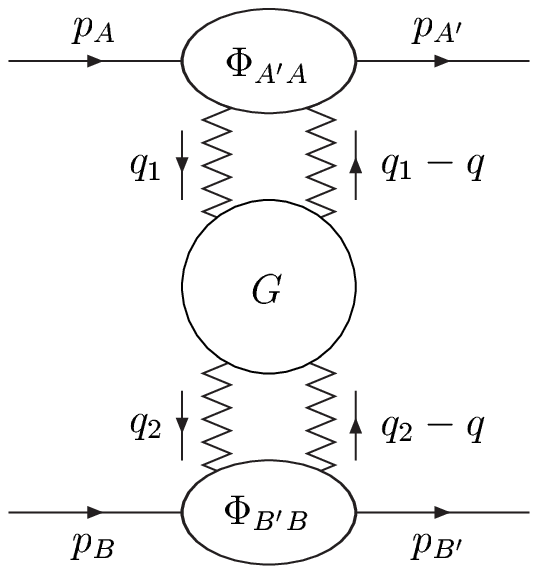}
\end{minipage}
\begin{minipage}{95mm}
\begin{flushright}
\includegraphics*[width=8cm]{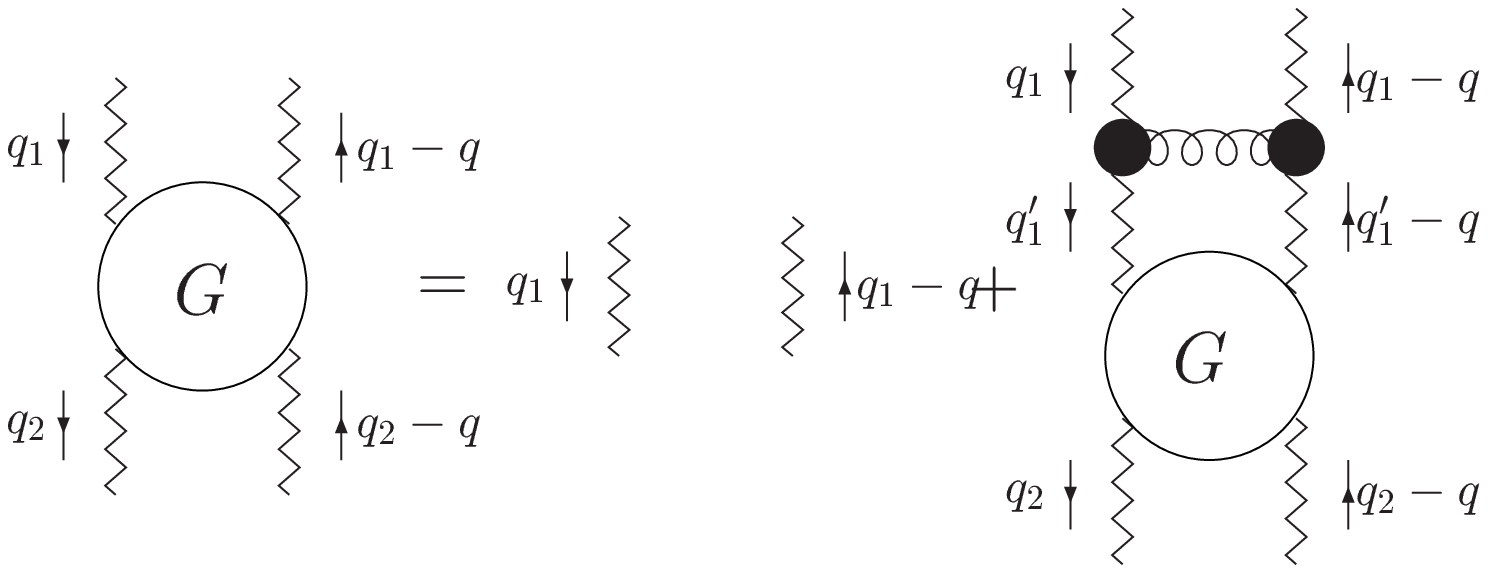}
\end{flushright}
\end{minipage}

\vspace{0.5cm}
\noindent {\scriptsize
Fig.~3. (Left) Diagrammatical representation of $A_{AB}^{A^{\prime }B^{\prime }}$ 
(for a definite color group representation) as derived from $s$-channel unitarity. 
The ovals are the impact factors of the particles $A$ and $B$, the circle is 
the Green's function for the Reggeon-Reggeon scattering.}

\vspace{0.3cm}
\noindent {\scriptsize
Fig.~4. (Right) Schematical representation of the BFKL equation in the LLA.}
\vspace{0.5cm}

\end{figure}

\section{BFKL in the LLA}
\label{LLA}

Amplitudes with quantum numbers in the $t$-channel different from the gluon ones
are obtained in the BFKL approach by means of unitarity relations, thus calling 
for inelastic amplitudes. In the LLA, the main contributions to the unitarity 
relations from inelastic amplitudes come from the multi-Regge kinematics, i.e. 
when rapidities of the produced particles are strongly ordered and their transverse 
momenta do not grow with $s$. In the multi-Regge kinematics, the real 
part\footnote{The imaginary part gives a next-to-next-to-leading contribution in 
the unitarity relations.} of the production amplitudes takes a simple factorized
form, due to gluon Reggeization,
\begin{equation}
A^{\tilde A  \tilde B +n}_{AB}=2s \Gamma_{\tilde A A}^{c_1} \!
\left(\prod_{i=1}^n \gamma _{c_ic_{i+1}}^{P_i}(q_i,q_{i+1})
\left(\frac{s_i}{s_R}\right) ^{\omega_i}\! \frac 1{t_i}\right) \!
\frac 1{t_{n+1}}\!
\left( \frac{s_{n+1}}{s_R}\right) ^{\omega_{n+1}}\!
\Gamma _{\tilde B B}^{c_{n+1}}\,,
\label{inelast_ampl}
\end{equation}
where $s_R$ is an energy scale, irrelevant in the LLA, 
$\gamma_{c_i c_{i+1}}^{P_i}(q_i,q_{i+1})$ is the (non-local) effective vertex 
for the production of the particles $P_i$ with momenta $k_i=q_i-q_{i+1}$ 
in the collisions of Reggeons with momenta $q_i$ and $-q_{i+1}$ and color indices
$c_i$ and $c_{i+1}$, $q_0\equiv p_A$, $q_{n+1}\equiv -p_B$, $s_i=(k_{i-1}+k_i)^2$,
$k_0 \equiv p_{\tilde A}$, $k_{n+1} \equiv p_{\tilde B}$ and $\omega_i$ stands 
for $\omega(t_i)$, with $t_i=q_i^2$. In the LLA, $P_i$ can 
be only the state of a single gluon (see Fig.~2). By using $s$-channel unitarity 
and the previous expression for the production amplitudes, the amplitude of the
elastic scattering process $A + B \longrightarrow A^\prime + B^\prime $ at high
energies can be written as
\begin{eqnarray}
A_{AB}^{A^{\prime }B^{\prime }} &=& \frac{is}{{\left( 2\pi \right) ^{D-1}}}
\int\frac{d^{D-2}q_{1}}{\vec{q}_{1}^{~2} \vec{q}_{1}^{\, \prime\, 2}}\int
\frac{d^{D-2}q_{2}}{\vec{q}_{2}^{~2} \vec{q}_{2}^{\, \prime \, 2}}
\int_{\delta-i\infty}^{\delta+i\infty} \frac{d\omega}{\mbox{sin}(\pi\omega)}
\sum_{R,\nu} \Phi _{A^{\prime}A}^{\left(R,\nu \right) }\left(
\vec{q}_{1};\vec{q};s_{0}\right)
\nonumber \\
&\times& \left[\left( \frac{-s}{s_{0}}\right)^{\omega}-
\tau\left(\frac{s}{s_{0}}\right)^{\omega}\right]
{G_{\omega }^{\left(R\right) }\left( \vec{q}_{1},\vec{q}_{2},\vec{q}\right)}
\Phi_{B^{\prime }B}^{\left(R,\nu \right)}\left(-\vec{q}_{2};-\vec{q};s_{0}\right)\,.
\label{elast_ampl_R}
\end{eqnarray}
Here and below $q_i^\prime \equiv q_i - q$, $q\sim q_\perp$ is the momentum 
transfer in the process, the sum is over the irreducible representations $R$
of the color group contained in the product of two
adjoint representations and over the states $\nu$ of these representations, 
$\tau$ is the signature equal to $+1 (-1)$ for symmetric (antisymmetric) 
representations and $s_{0}$ is an energy scale. $\Phi _{P^{\prime }P}^{\left(
R,\nu \right)}$ is the so-called impact factor in the $t$-channel color state 
$(R,\nu)$ and $G_\omega^{(R)}$ is the Mellin transform of the Green's functions for
Reggeon-Reggeon scattering (see Fig.~3). The dependence from $s$ is determined by 
$G_\omega^{(R)}$, which obeys the equation (see Fig.~4)
\begin{eqnarray}
\omega G_{\omega }^{(R)}(\vec{q}_{1},\vec{q}_{2},\vec{q}) &=& 
\vec{q}_{1}^{\:2}\vec{q}_{1}^{\:\prime\:2}
\delta^{(D-2)}(\vec{q}_{1}-\vec{q}_2) \nonumber \\
&+&\int \frac{d^{D-2}{q}_r}{\vec{q}_r^{\:2}\vec{q}_r^{\:\prime\:2}}
K^{(R)}(\vec{q}_{1},\vec{q}_r;\vec{q}) G_{\omega}^{(R)}(\vec{q}_r,\vec{q}_{2};
\vec{q})\;,
\label{BFKL}
\end{eqnarray}
whose integral kernel, 
\begin{equation}
K^{(R)}(\vec{q}_{1},\vec{q}_{2};\vec{q}) 
=[ \omega(-\vec{q}_{1}^{\:2})+\omega(-\vec{q}_{1}^{\:\prime\:2})]
\delta^{(D-2)}(\vec{q}_{1}-\vec{q}_{2}) 
+K_{r}^{(R)}(\vec{q}_{1},\vec{q}_{2};\vec{q})\;,
\label{kernel}
\end{equation}
is composed by a ``virtual'' part, related to the gluon trajectory, 
and by a ``real'' part, related to particle production in Reggeon-Reggeon 
collisions. In the LLA, the ``virtual'' part of the kernel takes contribution
from the gluon Regge trajectory with 1-loop accuracy, $\omega^{(1)}$, while
the ``real'' part takes contribution from the production
of one gluon in the Reggeon-Reggeon collision at Born level, 
$K_{RRG}^{(B)}$. The BFKL equation is Eq.~(\ref{BFKL}) specialized for 
$t=0$ and singlet quantum numbers in the $t$-channel. 

The representation~(\ref{elast_ampl_R}) of the elastic amplitude, 
$A + B \longrightarrow A^\prime + B^\prime$, derived from $s$-channel unitarity,
for the part with gluon quantum numbers in the $t$-channel ($R=8$, $\tau=-1$),
must reproduce the representation~(\ref{elast_ampl_octet}) with one Reggeized 
gluon exchange in the $t$-channel, with LLA accuracy. This consistency is called
``bootstrap'' and was checked in the LLA already in~\cite{BFKL}. Subsequently,
a rigorous proof of the gluon Reggeization in the LLA was constructed~\cite{BFL79}.

The part of the representation~(\ref{elast_ampl_R}) with vacuum quantum numbers
in the $t$-channel ($R=0$, $\tau=+1$) for the case of zero momentum transfer
is relevant for the total cross section of the scattering of particles $A$ and $B$,
via the optical theorem. In the LLA, it turns out that 
$\sigma_{AB}^{\mbox{\scriptsize LLA}} \sim s^{\omega_P^B}/\sqrt{\ln s}$, with
$\omega_P^B=4 \ln 2 N \, \alpha_s/\pi$. 
This relation shows that unitarity is violated, since 
the cross section overcomes the Froissart-Martin bound. This is obvious since,
in the LLA, only a definite set of intermediate states, as we have seen, contributes
to the $s$-channel unitarity relation. This means that the BFKL approach cannot be 
applied at asymptotically high energies. In order to identify the applicability
region of the BFKL approach, it is necessary to know the scale of 
$s$ and the argument of the running coupling constant, which are not
fixed in the LLA.

\section{BFKL in the NLA}
\label{NLA}

In the NLA, the Regge form of the elastic amplitude~(\ref{elast_ampl_octet})
and of the production amplitudes~(\ref{inelast_ampl}), implied by gluon
Reggeization, has been checked only in the first three orders of perturbation 
theory~\cite{FFKQ95-96}. In order to derive the BFKL equation in the NLA, gluon
Reggeization is assumed to be valid to all orders of perturbation theory. It becomes
important, therefore, to check the validity of this assumption. This will be done
at the end of this Section.

In the NLA it is necessary to include into the unitarity relations contributions
which differ from those in the LLA by having one additional power of $\alpha_s$
or one power less in $\ln s$. The first set of corrections is realized by
performing, only in one place, one of the following replacements in the production 
amplitudes~(\ref{inelast_ampl}) entering the $s$-channel 
unitarity relation:
$$
\omega^{(1)} \longrightarrow \omega^{(2)}\;,
\;\;\;\;\;\;\;\;\;\;
\Gamma_{P'P}^{c\;\mbox{\scriptsize (Born)}} \longrightarrow 
\Gamma_{P'P}^{c\;\mbox{\scriptsize (1-loop)}}\;,
\;\;\;\;\;\;\;\;\;\;
\gamma_{c_i c_{i+1}}^{G_i \mbox{\scriptsize (Born)}} \longrightarrow 
\gamma_{c_i c_{i+1}}^{G_i \mbox{\scriptsize (1-loop)}}\;.
$$
The second set of corrections consistes in allowing the production in the 
$s$-channel intermediate state of {\em one} pair of particles with rapidities 
of the same order of magnitude, both in the central or in the fragmentation 
region (quasi-multi-Regge kinematics). This implies one replacement among the 
following in the production amplitudes~(\ref{inelast_ampl}) entering the 
$s$-channel unitarity relation: 
$$
\Gamma_{P'P}^{c\;\mbox{\scriptsize (Born)}} \longrightarrow 
\Gamma_{\{f\} P}^{c\;\mbox{\scriptsize (Born)}}\;,
\;\;\;\;\;\;\;\;\;\;\;\;\;\;\;\;\;\;\;\;\;\;
\gamma_{c_i c_{i+1}}^{G_i \mbox{\scriptsize (Born)}} \longrightarrow 
\gamma_{c_i c_{i+1}}^{Q\overline Q \mbox{\scriptsize (Born)}}\;,
$$
$$
\gamma_{c_i c_{i+1}}^{G_i \mbox{\scriptsize (Born)}} \longrightarrow 
\gamma_{c_i c_{i+1}}^{GG \mbox{\scriptsize (Born)}}\;.
$$
Here $\Gamma_{\{f\} P}$ stands for the production of a state containing an extra
particle in the fragmentation region of the particle $P$ in the scattering
off the Reggeon, $\gamma_{c_i c_{i+1}}^{Q\overline Q \mbox{\scriptsize (Born)}}$
and $\gamma_{c_i c_{i+1}}^{GG \mbox{\scriptsize (Born)}}$ are
the effective vertices for the production of a quark anti-quark pair and
of a two-gluon pair, respectively, in the collision of two Reggeons. 

The detailed program of next-to-leading corrections to the BFKL equation
was formulated in Ref.~\cite{FL89}. They have been calculated over a period
of several years, mostly by one research group, lead by V.S.~Fadin (for an
exhaustive review, see Refs.~\cite{Fadin}). It turns out that also in the NLA
the amplitude for the high energy elastic process $A + B \longrightarrow A^\prime 
+ B^\prime$ can be represented as in Eq.~(\ref{elast_ampl_R}) and in Fig.~3. 
The Green's functions
obey an equation with the same form as Eq.~(\ref{BFKL}), with a kernel having
the same structure as in Eq.~(\ref{kernel}). Here the ``virtual'' part of the kernel
takes also the contribution from the gluon trajectory at 2-loop accuracy, 
$\omega^{(2)}$~\cite{FFKQ95-96}, while the ``real'' part of the kernel 
takes the additional contribution from one-gluon production in the Reggeon-Reggeon 
collisions at 1-loop order, $K_{RRG}^{(1)}$~\cite{FL93,FFQ94,FFK96,FFP01}, 
quark anti-quark pair production at Born level, 
$K_{RRQ\overline Q}^{(B)}$ (Ref.~\cite{FFFK97} for the forward case, 
Ref.~\cite{FFP99} for the non-forward case) and from two-gluon pair at 
Born level, $K_{RRGG}^{(B)}$ (Ref.~\cite{FKL97} for the forward case,
Ref.~\cite{FG00} for the non-forward, octet case). So, summarizing, the ``virtual'' 
part of the kernel is known in the NLA; as for the ``real'' part in the NLA, 
it is known completely for the singlet color representation in the $t$-channel 
in the forward case ($t=0$) and for the octet color representation in the 
$t$-channel in the non-forward case. The singlet NLA kernel in the non-forward 
case is not completely known, since the singlet $K_{RRGG}^{(B)}$ contribution 
is still missing.

The consistency between the representation~(\ref{elast_ampl_R}) 
of the elastic amplitude, $A + B \longrightarrow A^\prime + B^\prime$, derived 
from $s$-channel unitarity, for the part with gluon quantum numbers in the 
$t$-channel ($R=8$, $\tau=-1$), and the representation~(\ref{elast_ampl_octet}) 
with one Reggeized gluon exchange in the $t$-channel (``bootstrap'') is of crucial
importance in the NLA. In this approximation, indeed, gluon Reggeization was only 
assumed in order to derive the BFKL equation. Moreover, the check of the 
bootstrap is also a (partial) check of the correctness of calculations
which, as already pointed out, were performed mostly by one research group.
In the NLA, the bootstrap leads to two conditions to be verified~\cite{FF98},
one on the NLA octet kernel, the other on the NLA octet impact factors. The first
bootstrap condition has been verified at arbitrary space-time dimension for 
the part concerning the quark contribution to the kernel in massless 
QCD~\cite{FFP99}, while for the part concerning the gluon contribution to 
the kernel it has been verified in the $D \rightarrow 4$ limit~\cite{FFK00}.
The second bootstrap condition is process-dependent, therefore it should be 
checked in principle for every octet impact factor. So far,
it has been checked at arbitrary space-time dimension for quark and gluon 
NLA impact factors in QCD with massive quarks~\cite{FFKP00_G,FFKP00_Q}. 

As for the exchange of vacuum quantum numbers in the $t$-channel, the NLA
corrections to the BFKL kernel lead to a large correction to the BFKL Po\-me\-ron 
intercept~\cite{FL98,CC98}: $\omega_P= \omega_P^B (1- 2.4 \: \omega_P^B)$,
with $\omega_P^B = 4 \ln 2 N \alpha_s(\vec q^{\:2})/\pi$. A lot of papers 
have been devoted to the problem of this large correction (see, for 
instance,~\cite{NLA_Pomeron}). In my opinion, the BFKL Pomeron intercept itself has
not a special physical meaning. Instead, the BFKL approach can predict the full
amplitude of hard QCD processes in the NLA, as soon as NLA impact factors 
of colorless particles accessible to perturbative QCD are known in the 
next-to-leading-order. In this respect, one of the most interesting 
calculations, that of $\gamma^*\rightarrow\gamma^*$ impact factors in the NLA,
is in progress~\cite{FIK,BGQ}.

\section{Strong bootstrap}
\label{strong}

It has been proposed by Braun and Vacca~\cite{BV99} that
gluon Reggeization is realized also in the unphysical particle-Reggeon
scattering amplitude with gluon quantum numbers in the $t$-channel. This 
requirement leads to two so-called ``strong'' bootstrap conditions. One
of them fixes the process-dependence of the octet impact factors, stating that
any octet impact factor is proportional to the corresponding effective vertex
by a universal ``coefficient function''. The other states that this universal
coefficient function is ``eigenstate'' of the octet BFKL kernel, taken
as operator in the transverse space, with the gluon trajectory as ``eigenvalue''.
The coefficient function has been determined in the NLA by the known expression of 
the octet quark impact factors~\cite{FFKP00_Q} and has been used to check the
strong bootstrap condition on gluon impact factors~\cite{FFKP00} in the NLA. 
The strong bootstrap condition on the kernel is trivially satisfied in the LLA.
As for the NLA, it has been verified so far only for the quark 
part~\cite{BV99,FFKP00}. Although the physical role of the strong bootstrap
conditions is not yet clear, they are of practical importance, since their
fulfillment implies that of the ``soft'' ones~\cite{FF98}.

\end{document}